\documentclass[twocolumn,showpacs,preprintnumbers,amsmath,amssymb]{revtex4}

\usepackage{graphicx}
\usepackage{dcolumn}
\usepackage{bm}

\begin{document}

\preprint{APS/123-QED}

\title{B(E2) anomaly in $6^+$ isomers of $^{134-138}$Sn isotopes and neutron single-particle energies beyond N=82}
\author{Bhoomika Maheshwari}
\email{bhoomika.physics@gmail.com}
\author{Ashok Kumar Jain}%
\affiliation{Department of Physics, Indian Institute of Technology, Roorkee 247667, India.}
\date{\today}

\begin{abstract}
Isomeric studies in neutron-rich nuclei present a powerful tool to explore the structure at the nuclear extremes. We recently used the shell model calculations with Renormalized Charge Dependent Bonn (RCDB) effective interaction to calculate the properties of the $6^+$ seniority isomers in $^{134-138}$Sn in an attempt to resolve the anomalous B$(E2)$ behavior of the $6^+$ isomer in $^{136}$Sn [Phys. Rev. C 91, 024321 (2015)]. We further explore these isomers by using the generalized seniority scheme for multi-j orbitals recently presented by us [Phys. Lett. B 753, 122 (2016)]; the B$(E2)$ values so calculated reproduce the experimental data quite well, including the anomaly at $^{136}$Sn confirming the generalized seniority nature of the $6^+$ isomers. We then use the generalized seniority guided Large Scale Shell Model (LSSM) calculations, along with the latest single particle energies from Jones $\it{et}$ $\it{al.}$ [Nature (London) 465, 454 (2010)] and Allmond $\it{et}$ $\it{al.}$ [Phys. Rev. Lett. 112, 172701 (2014)] to estimate more accurate location of i$_{13/2}$ neutron orbital in the extreme neutron rich $N=82-126$ region. This entails a new sub-shell closure at $N=112$ due to the higher location of i$_{13/2}$ neutron orbital, also consistent with the choice of orbitals in the generalized seniority scheme. However, a small reduction in the f$_{7/2}$ two-body matrix elements is still required in the LSSM calculations to reproduce the experimental level energies as well as the transition probabilities in $^{134-138}$Sn isotopes in a consistent way.
\end{abstract}

\pacs{23.35.+g, 21.60.Cs, 23.20.Js, 27.60.+j}
\maketitle

\section{\label{sec:level1}Introduction}

Recent advancements in the experimental techniques have opened up new horizons in nuclear isomer studies near the nuclear limits, which obviously need a good theory to complement. Our recent atlas of nuclear isomers~\cite{jain} presents about $2450$ isomers across the chart of nuclides with several types of isomers and a rich variety of systematics. Nuclear isomers near the drip-lines have the potential to become a great tool to understand the detailed nuclear structural properties and the effective interactions in the region far from stability. One of the well known categories of the nuclear isomers is the “seniority isomers”, particularly near the magic configurations. The semi-magic Sn-isotopes have occupied a center stage because of the presence of the longest known chain of isotopes in the nuclear landscape extending from doubly magic $^{100}$Sn near the proton drip line to the doubly-magic $^{132}$Sn near the neutron drip line and beyond. 

The Sn-isotopic chain has now been extended beyond $^{132}$Sn in several remarkable experiments~\cite{zhang,korgul,beene,simpson}. More specifically, Simpson $\it{et}$ $\it{al.}$~\cite{simpson} recently presented measurements of $^{136-138}$Sn isotopes and brought out the anomalous behavior of $^{136}$Sn; they observed that the measured B$(E2)$ value for the decay of $6^+$ isomer in $^{136}$Sn is quite large as compared to the calculated values, both from the seniority as well as shell model calculations. They sought to resolve this anomalous behavior by reducing the diagonal and non-diagonal f$_{7/2}$ matrix elements by about $150$ keV. The resulting seniority mixing is then able to take care of the anomaly. 

We have also investigated~\cite{maheshwari1} this anomaly by using the RCDB effective interaction (also known as CWG interaction) in the shell model~\cite{brown1}. We were, however, able to explain the anomaly seen at $^{136}$Sn in terms of seniority mixing by reducing the diagonal and non-diagonal two-body matrix elements (TBME) of $\nu$f$_{7/2}^2$ just by $25$ keV in the RCDB interaction~\cite{maheshwari1}. It, however, confirmed that the TBME needs a reduction as pointed out by Simpson $\it{et}$ $\it{al.}$~\cite{simpson}.

In the present paper, we aim to study the $6^+$ isomers in $^{134-138}$Sn isotopes by using the generalized seniority scheme in a simple microscopic formalism recently presented by us~\cite{maheshwari2}. We find that the generalized seniority calculations for multi-j degenerate orbitals presented in this paper are able to reproduce the experimental B$(E2)$s as well as the said anomaly at $^{136}$Sn in a natural way. We also calculate the B$(E2)$ values for the first excited $2^+$ states and use the single measured value of B$(E2;2^+ \rightarrow 0^+)$ in $^{134}$Sn to predict the B$(E2; 2^+ \rightarrow 0^+)$ values in $^{136,138}$Sn, where no experimental data exist. We conclude that the n-rich $6^+$ Sn-isomers follow generalized seniority.

We also study these n-rich Sn-isomers within the shell model to check and validate the possible magic numbers, the realistic effective interactions and single particle states in the neutron-rich $N=82-126$ region. Recent measurements on the single particle states using the inverse kinematics by Jones $\it{et}$ $\it{al.}$~\cite{jones} and Allmond $\it{et}$ $\it{al.}$ ~\cite{allmond} give us reasonable information of the $N=82-126$ valence space which consists of 2f$_{7/2}$, 3p$_{3/2}$, 3p$_{1/2}$, 1h$_{9/2}$, 2f$_{5/2}$, and 1i$_{13/2}$ orbitals by spectroscopic studies of $^{133}$Sn; the single-particle energies for 2f$_{7/2}$, 3p$_{3/2}$, 3p$_{1/2}$ and 2f$_{5/2}$ orbitals have been determined. The location of 1h$_{9/2}$ has been tentatively assigned in these studies, while the placement of 1i$_{13/2}$ remained rather ambiguous. It may be noted that the earlier work on the single particle energies (SPE) of these orbitals in a decay study of fission fragments is due to Hoff $\it{et}$ $\it{al.}$~\cite{hoff}. These SPE provide the input to the calculations for the neighboring nuclei, which are still inaccessible in experiments. In the present paper, we employ  the LSSM calculations guided by the generalized seniority results to understand the $6^+$ isomers in n-rich nuclei. We try to validate the known single particle states and their energies, and also estimate the location of i$_{13/2}$ orbital. Our attempt is to reproduce the energy levels as well as the transition probabilities in a consistent way. While we are able to reproduce both the data quite well, we need to reduce the TBME of $\nu$f$_{7/2}^2$ by 25 keV along with a shift in the location of the i$_{13/2}$ orbital. This points towards the need for a pairing reduction in the n-rich nuclei beyond $^{132}$Sn, the doubly magic core, with an indication of a neutron-subshell gap at $112$. 

The present paper is divided into four sections. Section II briefly reports the theoretical framework involved in the generalized seniority calculations and LSSM calculations. We present the generalized seniority results in section III, where we compare them with the seniority guided LSSM calculations. We find the isomeric states in the n-rich nuclei may become a good handle to investigate the single particle energy (SPE) in the n-rich region. Section IV summarizes the present work.

\begin{figure}[!ht]
\includegraphics[width=9cm,height=8cm]{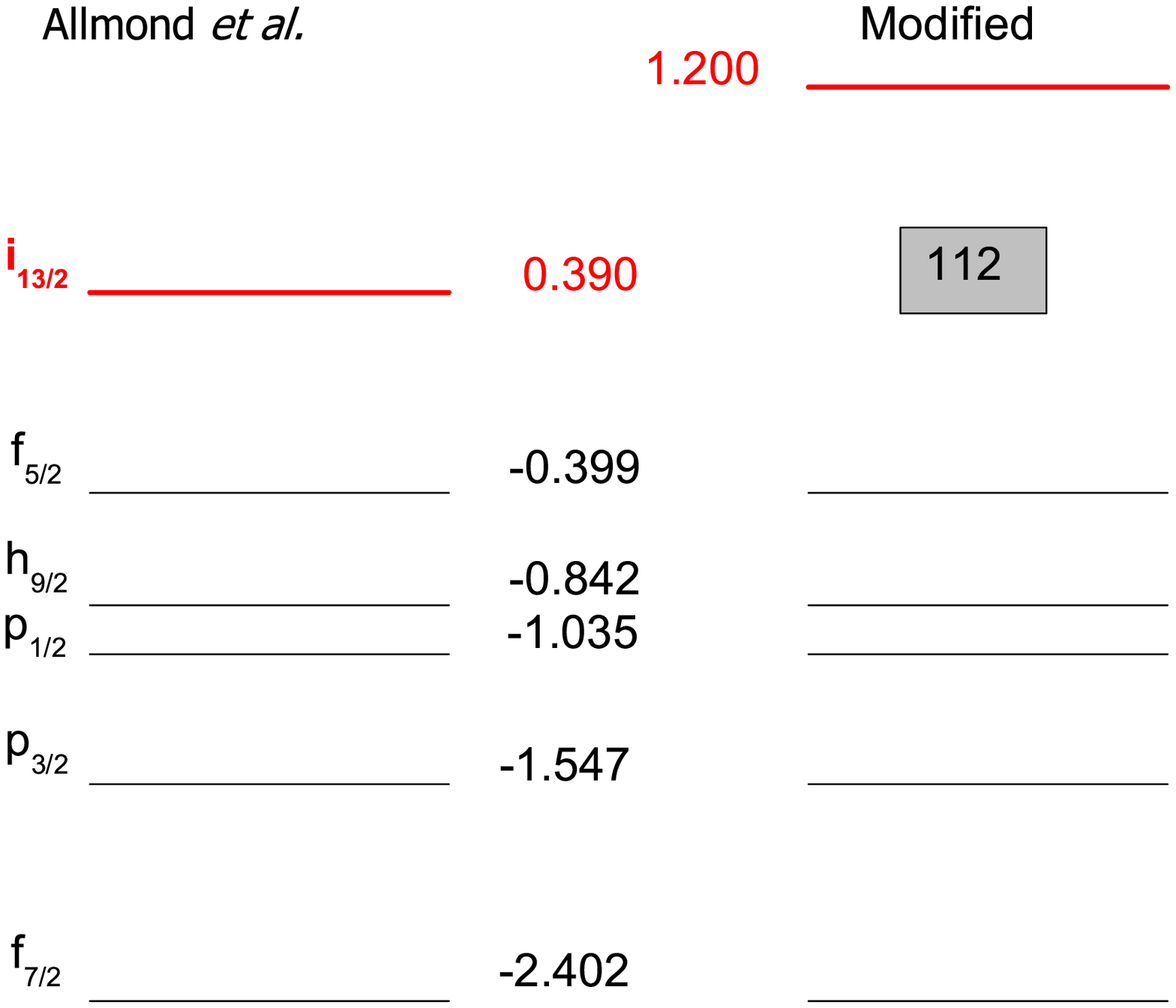}
\caption{\label{fig:spe}(Color online) Single particle energies of the orbitals in $N=82-126$ valence space, from Allmond $\it{et}$ $\it{al.}$~\cite{allmond} and the modified scheme used by us. }
\end{figure}

\begin{figure*}[!ht]
\includegraphics[width=16cm,height=10cm]{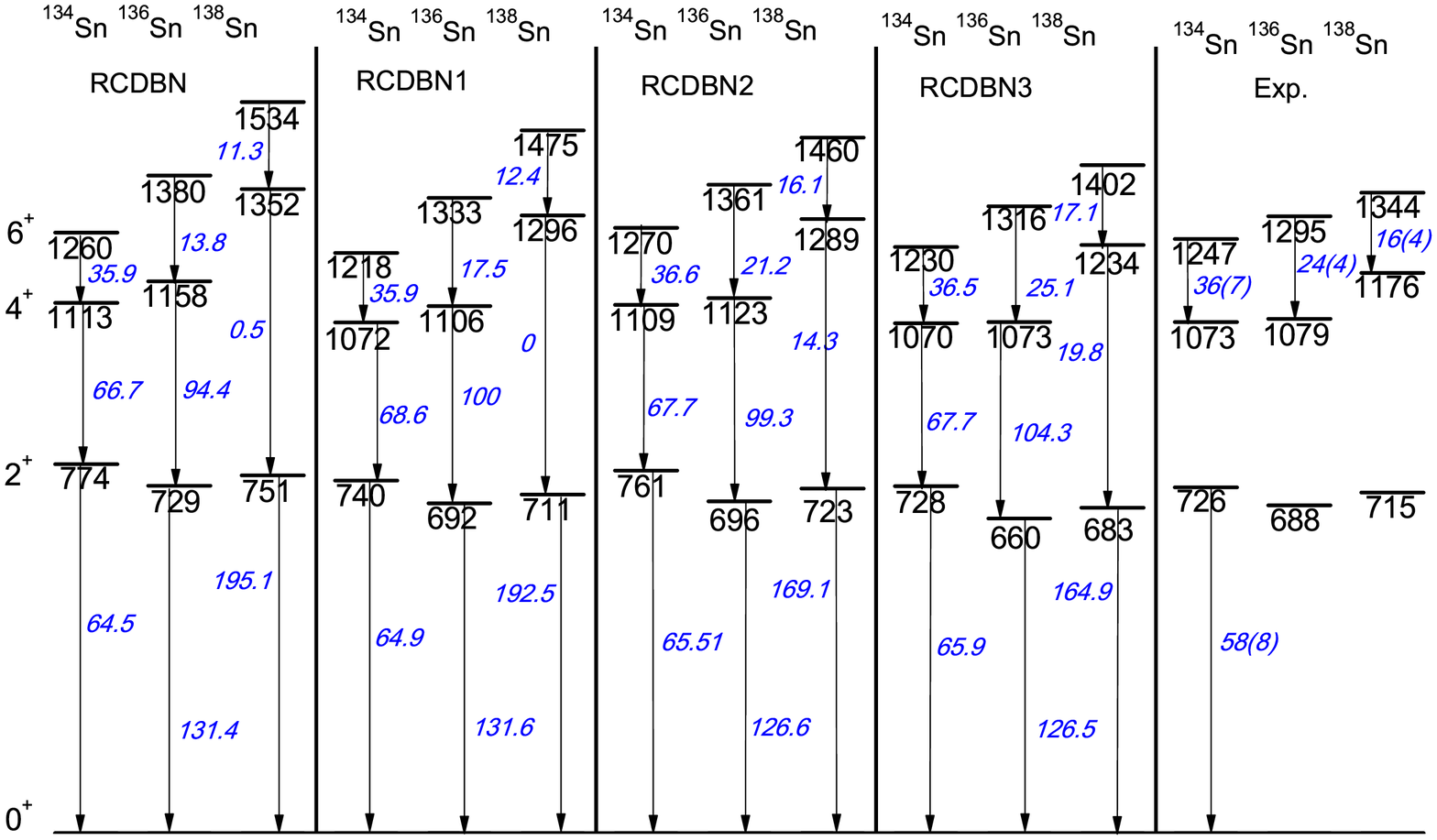}
\caption{\label{fig:level}(Color online) Level energies and B$(E2)$ values of the yrast $0^+$ to $6^+$ states in $^{134-138}$Sn isotopes by using RCDBN, RCDBN1, RCDBN2 and RCDBN3 interactions in LSSM calculations (first to forth panel from the left), in comparison with the experimental data (last panel). Experimental data are taken from~\cite{zhang,korgul,beene,simpson}. B$(E2)$ values, in units of $e^2 {fm}^4$, are listed in $italics$ along the transitions.}
\end{figure*}

\section{\label{sec:level2}Theoretical Framework}

\subsection{Generalized Seniority with many degenerate orbitals}

We have recently shown~\cite{maheshwari2} that the electric transition probabilities, for both even and odd $L$, behave similar to each other in multi-j environment and the generalized seniority scheme for degenerate orbitals. The B$(E2)$ between $J_i$ and $J_f$ states, hence, follow the relations similar to single-j case, simply by defining a mixed configuration as $\tilde{j} = j \otimes j'....$ along with the corresponding total pair degeneracy $\Omega= \frac{1}{2}(2 \tilde{j} +1)= \frac{1}{2} \sum \limits_j (2j+1)$ as
\begin{eqnarray}
B(E2)=\frac{1}{2J_i+1}| \langle \tilde{j}^n v l J_f || \sum_i r_i^2 Y^{2}(\theta_i,\phi_i) || \tilde{j}^n v' l' J_i \rangle |^2
\end{eqnarray}

This implies that the B$(E2)$ values show a parabolic behavior in the multi-j case, similar to the single-j case, depending upon the seniority of the states involved in the transition~\cite{maheshwari2,talmi}. The seniority reduction formula for the reduced matrix elements with seniority conserving $\Delta v=0$ and seniority changing $\Delta v=2$ transitions between the initial and final states, can be written as follows~\cite{maheshwari2}
\begin{eqnarray}
\langle \tilde{j}^n v l J_f ||\sum_i r_i^2 Y^{2}|| \tilde{j}^n v l' J_i \rangle = \Bigg[ \frac{\Omega-n}{\Omega-v} \Bigg] \langle \tilde{j}^v v l J_f \nonumber \\
||\sum_i r_i^2 Y^{2}|| \tilde{j}^v v l' J_i \rangle
\end{eqnarray}
\begin{eqnarray}
\langle \tilde{j}^n v l J_f || \sum_i r_i^2 Y^{2}||\tilde{j}^n v\pm 2 l' J_i \rangle  = 
\nonumber \\  \Bigg[ \sqrt{\frac{(n-v+2)(2\Omega+2-n-v)}{4(\Omega+1-v)}} \Bigg] \nonumber\\ \langle \tilde{j}^v v l J_f ||\sum_i r_i^2 Y^{2}|| \tilde{j}^v v\pm 2 l' J_i \rangle 
\end{eqnarray}

We use this formula in the following to calculate the B$(E2)$s, by allowing the mixing of all the orbitals except i$_{13/2}$. We find that the mixing of these orbitals is essential to understand the properties of the $6^+$ isomers as well as rest of the experimentally known yrast states in the n-rich Sn-isotopes, which were previously sought to be understood in terms of the f$_{7/2}$ orbital only. 

\begin{figure}[!ht]
\includegraphics[width=9cm,height=8.5cm]{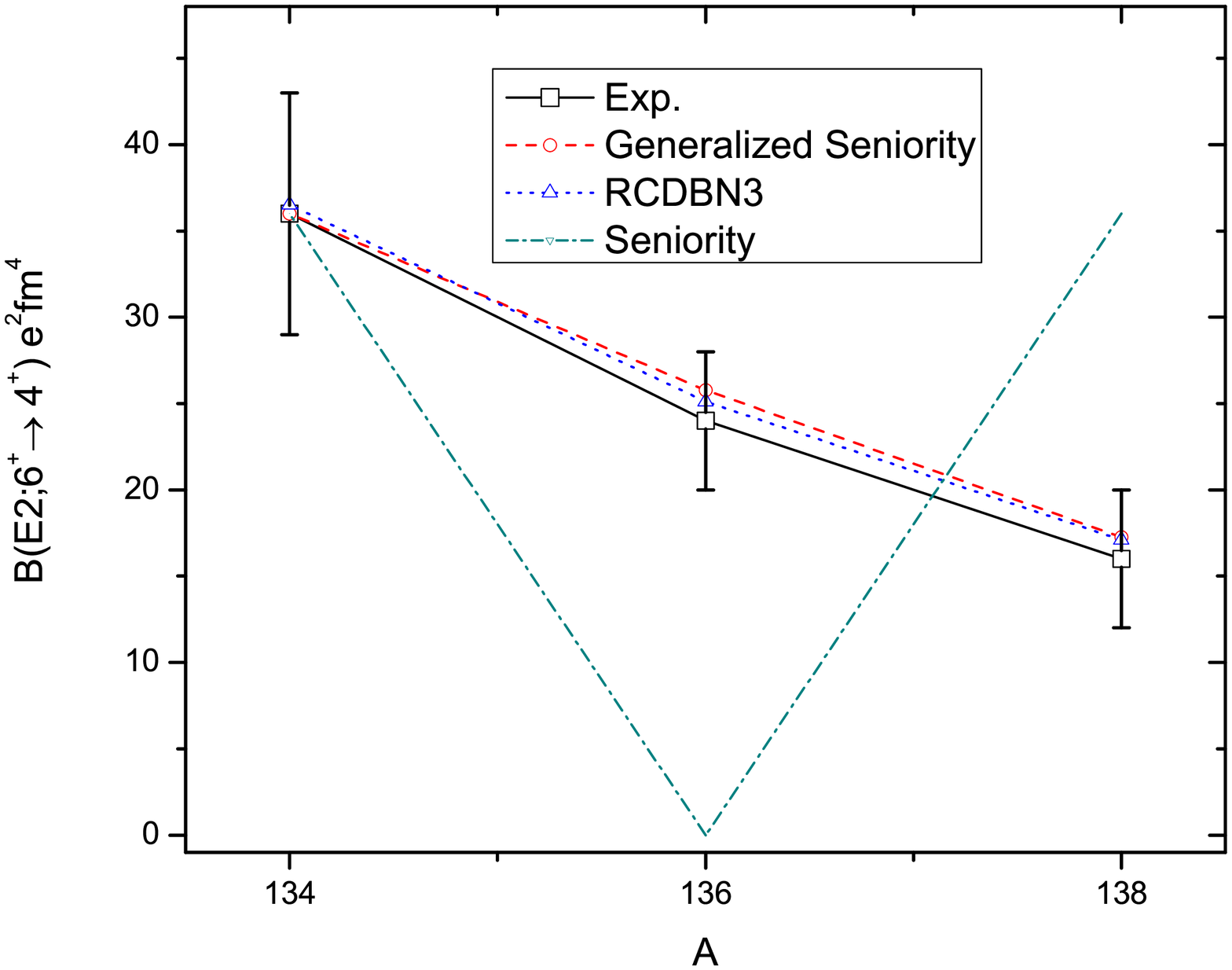}
\caption{\label{fig:be26}(Color online) B$(E2)$ variation of the $6^+$ isomeric states in $^{134-138}$Sn isotopes using the generalized seniority, LSSM calculations (RCDBN3) and seniority, respectively, in comparison with the known experimental data~\cite{zhang,korgul,beene,simpson}.}
\end{figure}

\begin{figure}[!ht]
\includegraphics[width=9cm,height=8.5cm]{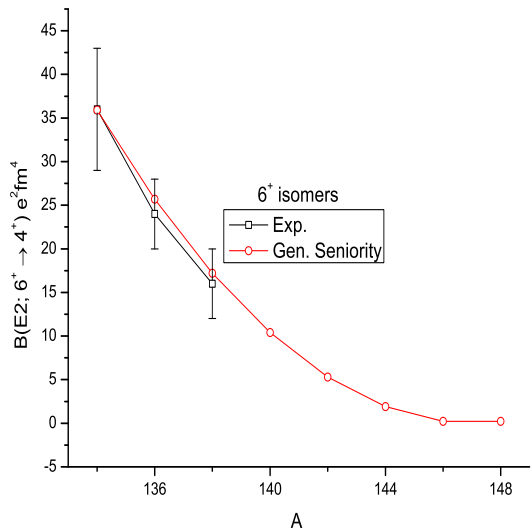}
\caption{\label{fig:prediction}(Color online) Predictions for B$(E2)$ values of the $6^+$ Sn-isomeric states using the generalized seniority up to $A=148$.}
\end{figure}

\begin{figure}[!ht]
\includegraphics[width=9cm,height=8.5cm]{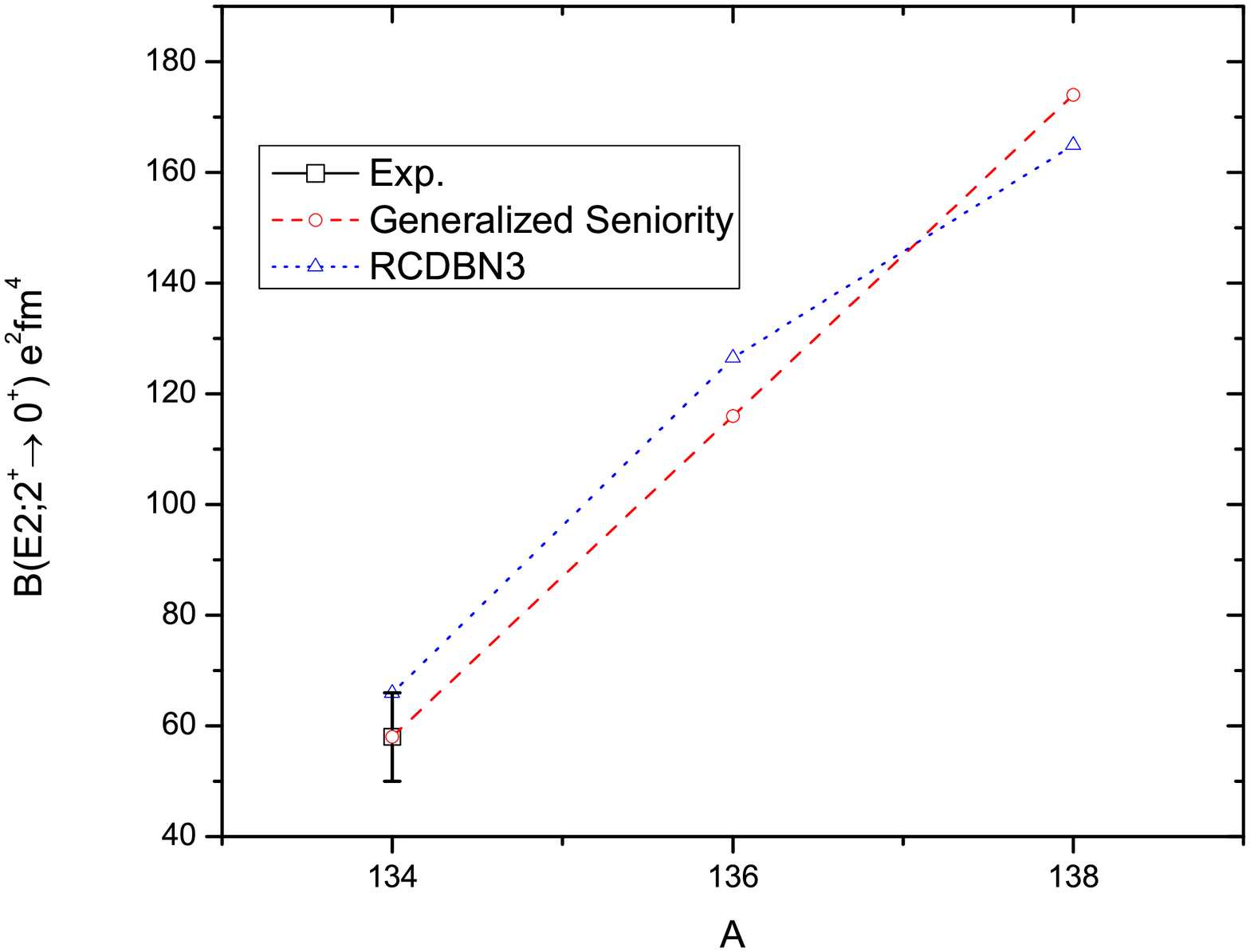}
\caption{\label{fig:be22} Same as Fig.~\ref{fig:be26}, but for the first excited $2^+$ states in $^{134-138}$Sn isotopes.}
\end{figure}

\subsection{LSSM calculations}

We also present the generalized seniority guided LSSM calculations for $^{134-138}$Sn isotopes, by using the latest SPE measurements of Jones $\it{et}$ $\it{al.}$~\cite{jones}, and Allmond $\it{et}$ $\it{al.}$~\cite{allmond}, which report the SPE for the orbitals in the $82-126$ neutron valence space. These valence orbitals are 2f$_{7/2}$, 3p$_{3/2}$, 3p$_{1/2}$, 1h$_{9/2}$, 2f$_{5/2}$, and 1i$_{13/2}$, where the location of i$_{13/2}$ remains quite uncertain. We have used Nushell code~\cite{brown2} along with the new RCDB interaction~\cite{brown1} for LSSM calculations, which we have named as RCDBN, wherein all the single particle energies as quoted by Allmond $\it{et}$ $\it{al.}$~\cite{allmond} have been used to calculate the B$(E2)$s in the n-rich Sn-isotopes. The TBMEs of the interaction have not been changed. This interaction assumes the doubly-magic $^{132}$Sn as an inert core. The harmonic oscillator potential has been chosen with the parameter $\hbar \omega = 45A^{-1/3}-25A^{-2/3}$. The SPE for these neutron orbitals are shown in the left panel of Fig.~\ref{fig:spe}. We have used the effective neutron charge as $0.65$, same as used in the previous paper~\cite{maheshwari1}. The calculated results reproduce the measured data for $^{134}$Sn quite well, but not for the other two $^{136,138}$Sn isotopes, due to the anomaly discussed earlier~\cite{simpson,maheshwari1}. 

\section{\label{sec:level3}Results and Discussion}

In our previous paper~\cite{maheshwari1}, we used the prescription of Simpson $\it{et}$ $\it{al.}$~\cite{simpson}, and were able to explain the anomaly at $^{136}$Sn by reducing the diagonal and non-diagonal $\nu$f$_{7/2}^2$ TBME by $25$ keV. In the present work, we also focus on the location of i$_{13/2}$, and carry out the LSSM calculations by shifting its position to $1.2$ MeV (original value $0.39$ MeV), now named as RCDBN1 (see Fig.~\ref{fig:spe} for the modified scheme). We find that the calculated results from RCDBN1 still need an improvement to explain the experimental data. In the second calculation, we reduce the diagonal and non-diagonal two-body matrix elements (TBME) of $\nu$f$_{7/2}^2$ configuration by $25$ keV while keeping the i$_{13/2}$ orbital unchanged at $0.39$ MeV (RCDBN2) with significant improvement in the  results. However, the level energies still need an improvement. In the third calculation, we shift the position of i$_{13/2}$ orbital to $1.2$ MeV and also reduce the TBME by $25$ keV (RCDBN3). We find that the calculations by using the modified interaction RCDBN3 reproduce the level energies as well as the B$(E2)$ values in a consistent way. We may note that a large number of calculations were done to check the best location of the i$_{13/2}$ orbital. We find that placing it below as well as above $1.2$ MeV deteriorates the results. 

We have plotted the calculated level energies for the yrast $0^+$ to $6^+$ states of $^{134-138}$Sn isotopes in Fig.~\ref{fig:level}. The graph has been divided into five different panels, where the first panel presents the results from the RCDBN interaction. The next three panels present the results from the modified interactions, RCDBN1, RCDBN2 and RCDBN3, respectively, and the last panel exhibits the experimental data for comparison. The figure also lists the measured and calculated B$(E2)$ values in units of $e^2{fm}^4$ for all the transitions in $italics$ (blue color online).

To sum up, the RCDBN interaction is unable to reproduce the measured B$(E2)$ values for the $6^+$ isomers, particularly in $^{136,138}$Sn, and leads to the same anomaly as discussed earlier~\cite{simpson,maheshwari1}. As we modify the interaction by shifting the SPE for the i$_{13/2}$ orbital in the modified interaction RCDBN1, we find that the calculated values come closer to the measured ones. The calculations having unchanged location of i$_{13/2}$ orbital, but with the reduced TBME denoted as RCDBN2, also improve the transition probabilities but level energies are overestimated. Therefore, we move to the combination of both the possibilities, and get the best fit with a shift in the position of i$_{13/2}$ orbital and a $25$ keV reduction of TBME as defined in the RCDBN3 interaction. The calculated values with this third modified interaction reproduce the measured values, both the level energies and the transition probabilities, quite well. The level energies from RCDBN3 show the best fit to the measured data for the yrast $0^+$ to $6^+$ states in all the three even-even isotopes. Note that the i$_{13/2}$ orbital is shifted to $1.2$ MeV and the f$_{7/2}$ TBME are also reduced by $25$ keV in the RCDBN3 interaction. This strongly suggests that the location of i$_{13/2}$ orbital needs to be shifted in a significant way to get the best fit of both the level energies and transition probabilities. This results in a sub-shell gap at $N=112$, though a TBME reduction is still required in realistic effective interaction for the n-rich region beyond $^{132}$Sn.

We then plot the generalized seniority results for the B$(E2)$ values along with the experimental data for the $6^+$ isomeric states in Fig.~\ref{fig:be26}, where one can see that these calculations reproduce the measured data quite well for all the three isotopes. Note that these calculations have been done by using all the orbitals in the $N=82-126$ valence space except i$_{13/2}$. The pair degeneracy $\Omega$ value is calculated as $15$ for this case. The B$(E2)$ values have been calculated by fitting the measured value of $^{134}$Sn.

We find that the generalized seniority results remove the anomaly at $^{136}$Sn in a quite natural way, whereas pure seniority result (using f$_{7/2}$ only) falls quite far from the experimental data. We also present the LSSM results from RCDBN3 interaction in Fig.~\ref{fig:be26} for comparison, which are in line with the generalized seniority results. Both the approaches reproduce the experimental data quite well, for the $6^+$ isomers in these nuclei. In Fig.~\ref{fig:prediction}, we present the predictions of B$(E2)$ values for $6^+$ isomers up to $A=148$. We can see that the B$(E2)$ values will continue to decrease with mass number; however, we do not expect all the isotopes to be within reach of experiments as the n-drip line is only tentatively known.  

We also present the B$(E2;2^+ \rightarrow 0^+)$ values in Fig.~\ref{fig:be22}, both from the generalized seniority and the LSSM calculations. We find that results from both the approaches follow each other. Note that only one measured value is available for B$(E2;2^+ \rightarrow 0^+)$ at $^{134}$Sn, while no measurements are available for the B$(E2;4^+ \rightarrow 2^+)$ values. This strongly suggests the need of measurements for these values, which may be able to settle this point further. 

We, therefore, confirm the location of the i$_{13/2}$ neutron orbital to be about $1.2$ MeV. This strongly suggests that the i$_{13/2}$ orbital is far from the other five orbitals of this valence space, and leads to a sub-shell gap within the valence space. This also points towards lesser mixing of i$_{13/2}$ orbital with other orbitals in this valence space supporting the choice of orbitals in generalized seniority. Hence, we may conclude that the observed anomaly in B$(E2)$ value at $^{136}$Sn may be understood simply by following the generalized seniority scheme. No core excitations are needed for this, which further confirms the rigidity of $N=82$ shell closure~\cite{maheshwari1}. Also, $N=90$ may not be the next sub-shell closure (obtained by filling of f$_{7/2}$ only), though we do find a gap at $N=112$ between the i$_{13/2}$ orbital and rest of the orbitals in this valence space~\cite{sarkar}. 

\section{Conclusions}

To conclude, we have used the generalized seniority scheme to study the $6^+$ isomers in $^{134-138}$Sn isotopes. It reproduces the experimental trend quite well and explains the B$(E2)$ anomaly of these isomers in a quite natural way. The success of this scheme in explaining the available experimental data suggests that the states in these n-rich isotopes are actually generalized seniority states rather than pure seniority ones, and points towards a certain gap between i$_{13/2}$ and the remaining orbitals of the valence space. We also study the first excited $2^+$ states within the generalized seniority scheme. This scheme also helps in predicting some spectroscopic numbers near the nuclear limits. To validate the configuration prescribed by the generalized seniority and check the role of SPE, particulalrly the tentatively assigned i$_{13/2}$ orbital energy, we present the results of the generalized seniority guided LSSM calculations. These results strongly point to a higher position of the i$_{13/2}$ orbital at $1.2$ MeV, which reproduces the measured values, both the level energies and the transition probabilities quite well, with a $25$ keV reduction in the diagonal and non-diagonal TBME of $\nu$f$_{7/2}^2$. We conclude from these results that the $N=82$ magic number remains quite robust even in the extreme n-rich region. We also obtain a partial gap at $N=112$, which is consistent with the generalized seniority results. 

\begin{acknowledgments}
One of the authors (BM) gratefully acknowledges the financial support in the form of Senior Research Fellowship from Ministry of Human Resource Development (Government of India).
\end{acknowledgments}

\newpage 
\bibliography{apssamp}

\end{document}